Scaling of Relativistic Shear Flows with Bulk Lorentz Factor

Edison Liang[1], Wen Fu[1], Markus Böttcher[2], Parisa Roustazadeh[2]

[1] Rice University, Houston, TX 77005

[2] North-West University, Potchefstroom, 2520, South Africa

**Abstract**

We compare Particle-in-Cell simulation results of relativistic electron-ion shear flows with different bulk Lorentz factors, and discuss their implications for spine-sheath models of blazar versus gamma-ray burst (GRB) jets. Specifically, we find that most properties of the shear boundary layer scale with the bulk Lorentz factor: the lower the Lorentz factor, the thinner the boundary layer, and the weaker the self-generated fields. Similarly, the energized electron spectrum peaks at an energy near the ion drift energy, which increases with bulk Lorentz factor, and the beaming of the accelerated electrons gets narrower with increasing Lorentz factor. This predicts a strong correlation between emitted photon energy, angular beaming and temporal variability with the bulk Lorentz factor. Observationally, we expect systematic differences between the high-energy emissions of blazars and GRB jets.

Subject Keywords: Shear Flow; Gamma-Ray Bursts; Quasars

## 1. INTRODUCTION

Unveiling the composition of relativistic jets of active galactic nuclei (AGN) and gamma-ray bursts (GRB), and the mechanisms of particle acceleration to ultrarelativistic energies within these jets, is among the prime outstanding issues in gamma-ray astronomy, as probed by the Fermi Gamma-Ray Space Telescope and ground-based Atmospheric Cherenkov Telescopes, such as H.E.S.S., MAGIC, and VERITAS, and the future Cherenkov Telescope Array (CTA). The physics of relativistic jets of AGN is most directly probed by observations of blazars, whose jets are oriented at a small angle with respect to our line of sight. Their broad-band nonthermal continuum emission consists of two broad emission components, and is almost certainly produced in small, localized regions within the relativistic jet. It is commonly accepted that the radio through optical/UV (and in some cases X-ray) emission from blazars is synchrotron emission from relativistic particles. Leptonic models for the high-energy emission of blazars propose that the X-rays and gamma-rays from blazars are the result of Compton upscattering of lower-energy photons by the same relativistic electrons (see, e.g. Boettcher 2007 for a review of blazar emission models).

There are several lines of evidence which suggest that the jets in blazars exhibit at least a two-component structure: a mildly relativistic, outer sheath with higher density, carries most of the kinetic energy of the jet, while a fast, highly relativistic inner spine of low co-moving particle density carries most of the angular momentum. Direct observational evidence for radially structured spine-sheath jets comes from the limb-brightening of blazar and radio galaxy jets revealed in VLBI observations (Giroletti 2004). Prompted by such evidence, Ghisellini (2005) proposed the radiative interaction between a fast, inner spine and a slower sheath in a blazar jet as a way to overcome problems with extreme bulk Lorentz factors required by spectral fits to several TeV BL Lac objects. Hydrodynamic/MHD simulations of spine-sheath jets (Meliani & Keppens 2007, 2009, Mizuno 2007) indicate that the sheath, in combination with a poloidal magnetic field, aids in stabilizing the jet. Although Kelvin-Helmholtz-type instabilities (KHI,

Chandrasekhar 1981) may develop at the spine-sheath interface and lead to turbulent mixing of the two phases, they may not disrupt the jet out to large distances from the central engine (Meliani & Keppens 2007, 2009). The MHD turbulence developing at the spine-sheath interface of relativistic jets (Zhang et al 2009) offers a promising avenue for relativistic particle acceleration in radio-loud AGNs and GRBs. However, the MHD approximation cannot directly address the creation of magnetic fields from unmagnetized shear flows or the acceleration of nonthermal particles.

The kinetic physics of relativistic shear flows has been successfully simulated using Particle-in-Cell (PIC, Birdshall & Langdon 1991) simulations (Alves et al 2012, 2014, 2015, Grismayer et al 2013, Liang et al 2013ab, Nishikawa et al 2013, 2014, 2016). In our previous papers (Liang et al 2013b, 2016), we have shown that ion-dominated relativistic shear flows lead to the creation of ordered dc electromagnetic (EM) fields near the shear boundary via the electron counter-current instability (ECCI), and the development of highly relativistic electron distributions peaking near the ion kinetic energy. However, those simulations assumed a high spine Lorentz factor ($\Gamma = 451$ in the central engine frame, $p_o=15$ in the center-of-momentum (CM) frame). Hence those results are more relevant to GRBs (Liang 2013b, 2016) than to AGNs.

In this paper, we present new PIC simulation results for a more moderate bulk Lorentz factor ($p_o=5$, $\Gamma=51$), relevant to radio-loud AGN, in particular blazars, in which bulk Lorentz factors $\Gamma \sim O(10)$ are typically inferred from superluminal motion and radio brightness-temperature arguments (Jorstad 2005, Hovatta 2009). We will systematically compare the $p_o=5$ shear boundary with the $p_o$=15 shear boundary. To simplify the comparison, we first focus on pure electron-ion (e-ion) plasmas. Generalization to mixtures of e-ion and electron-positron (e+e-ion) plasmas does not alter our major conclusions, and will be briefly mentioned at the end of Sec.2.

## 2. COMPARISON OF $p_o$=5 AND $p_o$=15 SHEAR BOUNDARIES.

As in our previous shear flow PIC simulations (Liang et al 2013ab, 2016), we use the 2.5D (2D space, 3-momenta) code Zohar-II (Birdsall & Langdon 1991, Langdon & Lasinski 1976) as the primary simulation tool. Though our Zohar-II simulation box is limited to 1024x2048 cells, this code has high numerical fidelity, and the numerical Cerenkov instability (NCI, Godfrey 1974, 1975) is strongly suppressed (Godfrey and Langdon 1976). Hence it is well suited for simulations with relativistic particle drifts. In all ion-dominated shear flows, the T-mode (Liang et al 2013a) in the y-z plane (Fig.1) saturates at very low amplitude compared to the P-mode (Liang et al 2013a) in the x-y plane, and has negligible effects on the shear boundary structure (Liang et al 2013b, confirmed by both 2.5D runs in the y-z plane and 3D runs). Hence we focus on the 2D P-mode (Fig.1) results in the x-y plane in this paper. All simulations are performed in the CM frame with periodic boundary conditions and initial temperature kT = 2.5 keV for both electrons and ions ($m_i/m_e$=1836). *Throughout this paper and in all figures, distances are measured in units of electron skin depth $c/\omega_e$ ($\omega_e$ = electron plasma frequency) and times are measured in units of $1/\omega_e$.* We normalize the initial density n=1 so that the cell size = $c/\omega_e$. The plasmas are initially unmagnetized. Initially right-moving plasma occupies the central 50% of the y-grid (hereafter called the "spine"), while initially left-moving plasma occupies the top 25% and bottom 25% of the y-grid (hereafter called the "sheath")(Fig.1). To increase numerical stability, we used time-step $\Delta t = 0.1/\omega_e$. Overall energy conservation was better than 1%.

We first compare the main features of $p_o=5$ and $p_o=15$ shear boundaries. Figure 2 shows the energy flows between ions, electrons and EM fields for the two runs. We see that in both cases, the electron and ion energies reach equipartition after $t\omega_e \sim 9000$, and EM energy saturates at ~ 12% of total energy, showing that the e-ion equipartition and EM energy saturation are insensitive to $p_o$. Figure 3 compares the spatial profiles of $\mathbf{B_z}$, $\mathbf{E_y}$, $\mathbf{E_x}$, $\mathbf{J_x}$, and net charge $\rho = (n_+-n_-)$ at $t\omega_e=1000$, 3000 and 12000 respectively for the two runs. While the overall patterns are qualitatively similar, the shear boundary layers of the $p_o=5$ case are thinner than those of the $p_o=15$ case by ~ factor of two. This is not unexpected since the thickness of the boundary layer should be related to the relativistic skin depth and relativistic gyroradius, both of which increase with increasing $p_o$. The maximum values of the dc fields ($\mathbf{B_z}$, $\mathbf{E_y}$) are also lower for $p_o=5$ than $p_o=15$ (Fig.3ab). Figure 4 compares the x-averaged density profiles of ions, electrons and net charge as functions of y for the two runs. This shows that the ion vacuum gap created by magnetic expulsion from the shear interface is present in both runs, but the gap is wider for $p_o=15$ than for $p_o=5$ due to stronger dc fields. This robust ion vacuum gap is a unique feature of relativistic ion-dominated shear flows, which sustains the separation of the opposing flows and the long-term stability of the laminar boundary layer structure against turbulent mixing of opposing ions. Electrons are evacuated less than the ions due to their mobility, leading to charge separation and the formation of a triple layer (double capacitor) at the shear boundary and associated $\mathbf{E_y}$ fields (cf. Fig.3b). Inductive $\mathbf{E_x}$ fields are generated adjacent to the boundary layer by $\partial \mathbf{B_z}/\partial t$ (Fig.3c), which accelerates the electrons and decelerates the ions.

Figure 5 compares the electron and ion energy distributions for the two runs at late times. Because the bulk of particle acceleration/deceleration is done by the $\mathbf{E}_x$ fields, the artificial periodic y-boundary condition turns out to have little effect on the late-time electron and ion distributions, as we had previously demonstrated using much larger y-grids (Liang 2013b, 2016). In both runs the electron spectrum exhibits a narrow peak near the (decelerated) ion drift kinetic energy. In the $p_o$=5 case, the electron spectrum peaks at $\gamma_e$~3000, consistent with the ion energy peak at ~$2.5 m_i c^2$ (hence kinetic energy ~$1.5 m_i c^2$). Similarly, for the $p_o$=15 case, the electron spectrum peaks at $\gamma_e$~14000, consistent with the ion energy peak at ~$7 m_i c^2$ (Liang et al 2013b). This confirms the scaling of the electron peak energy $\gamma_e$ with $p_o$. As we discuss below in Sec.3, in the context of synchrotron models, the electron peak energy $\gamma_e$ can be related to the synchrotron critical frequency (Rybicki & Lightman 1979) via $\omega_{cr} \sim \gamma_e^2 \omega_B$, where $\omega_B = eB/mc$ is the electron gyrofrequency (=Lamor frequency). On the other hand, for Compton models, the inverse Compton peak is located at $\omega_{IC} \sim \gamma_e^2 \omega_o$, where $\omega_o$ is the characteristic soft photon energy ($\omega_o \sim \omega_{cr}$ for SSC models, Boettcher 2007). Even though pure e-ion shear flows do not accelerate electrons much above the ion kinetic energy (Fig.5), when we add a moderate amount of e+e- plasma into the e-ion plasma, a power-law tail eventually develops above $\gamma_e$, due to the presence of nonlinear EM waves created outside the dc slab fields of Fig.3 (Liang et al 2013b), which scatter the leptons stochastically to form the power-law tail. We observe power-law tails develop in both the $p_o$=5 and $p_o$=15 cases, but preliminary results suggest that the power-law slope may vary with both $p_o$ and e+/ion ratio. Details remain to be investigated systematically.

## 3. APPLICATIONS TO BLAZARS AND GRBS

Assuming that blazar and GRB jets indeed have a spine-sheath structure, our shear boundary PIC simulations results above should be applicable to the local emission properties of the spine-sheath interface. To better visualize the differences in particle momentum distribution and radiation characteristics between the $p_o=5$ and $p_o=15$ cases, it is better to Lorentz boost the particle momenta from the CM frame of Sec.2 back to the *"laboratory" frame (LF) in which the sheath is initially at rest, and the spine moves with the bulk Lorentz factor $\Gamma=2p_o^2+1$*. Figures 6 & 7 compare various phase plots for the two runs, after Lorentz boosting (in the **–x** direction) from the CM frame back to the LF. We see that for $p_o=5$ (Figs.6a, 7a), spine electrons are accelerated to peak at $\gamma_{Lab} \sim 30000$ or 15 GeV, whereas for $p_o=15$ (Figs.6b,7b), spine electrons are accelerated to peak at $\gamma_{Lab} \sim 4.4 \times 10^5$ or 220 GeV. The highest-energy spine electron momenta achieve more extreme anisotropy ($\mathbf{p_{xLab}} \ggg \mathbf{p_y}$) for $p_o=15$ and than for $p_o=5$, while the beam angle $|\mathbf{p_y}/\mathbf{p_{xLab}}|$ decreases exponentially with increasing energy for both $p_o=5$ and $p_o=15$ (Fig.8). In fact, on average both beam angles are much narrower than simple Doppler boosting of an isotropic distribution in the spine rest frame to the LF ($1/\Gamma$, red dashed line). Observationally, we therefore expect GRB jets to emit much harder radiation with narrower beaming and more rapid time variability than blazar jets, and the photon energy should be correlated with time variability and anti-correlated with beam angle. Such observational predictions should be testable.

In summary, our PIC simulation results show that efficient lepton acceleration up to $\gamma_e \sim \Gamma_i$ $m_i/m_e$ occurs in relativistic shear boundary layers, and proceeds in a strongly anisotropic manner. The highest-energy leptons are beamed into an angle much narrower than $1/\Gamma$ in the laboratory frame. In the process of Compton scattering by relativistic leptons, the scattered, high-energy photon emerges in the direction of the scattering lepton. Hence jets viewed in the direction tangential to the shear boundary will exhibit very hard radiation spectra, much beyond the usual spectral hardening effect due to bulk Doppler boosting of a co-moving isotropic particle distribution (which is just a shift of the peak frequency by factor $\Gamma$). On the other hand, jets viewed at substantial off-axis angles (assuming that the shear layer is largely parallel to the global jet axis) will exhibit softer spectra. These beaming effects should become more acute for GRBs (Meszaros 2002, Piran 2004, Preece et al 1998) than blazars, and more extreme for the Compton peak than the synchrotron peak. The narrow beaming may also explain the minute-scale rapid time-variability of some blazars (Tavecchio and Ghisellini, private communications).

The results presented above indicate that relativistic shear layers in ion-dominated plasmas are capable of producing relativistic electron distributions in the CM frame with pronounced peaks at $\gamma_e \sim$ few x$10^3$ for blazar Lorentz factors $\Gamma \sim 10$. In the presence of a magnetic field of B = $B_G$ Gauss in the CM frame, this results in an observed synchrotron peak frequency of $\nu \sim 10^{14} B_G$ Hz in the LF, typically observed in low-synchrotron-peaked (LSP) blazars, i.e. flat-spectrum radio quasars and low-frequency-peaked BL Lac objects.

These same electrons will then also produce gamma-rays via Compton up-scattering of the co-spatially produced synchrotron photons (the synchrotron self-Compton (SSC) process) and possibly photons produced external to the jet (the external-Compton (EC) process), e.g., in the broad-line region or infrared-emitting dusty torus around the central accretion flow (Boettcher 2013). Synchrotron photons can be up-scattered (SSC) in the Thomson regime, which is expected to be the case for blazars for any plausible magnetic-field value. This will then result in a peak photon energy of the SSC emission of $\sim$ few MeV $B_G$. Gamma-ray emission of LSP blazars is often dominated by SSC emission (Boettcher 2013). External photons with stationary-frame energy $h\nu_o \sim$ eV will be Compton up-scattered in the Klein-Nishina regime to yield maximum observed photon energies of $\sim$ 15 GeV in LF. This is consistent with the gamma-ray peaks in LSP blazars typically being located at 100 MeV to GeV.

These estimates illustrate that for characteristic values of bulk Lorentz factor $\Gamma \sim 10$, the shear boundary energization scenario predicts synchrotron peaks in the IR-optical and EC gamma-ray peaks up to the GeV regime, as typically observed in LSP blazars (Abdo 2010), along with SSC-dominated hard X-ray and soft gamma-ray emission, peaking around $\sim$ few MeV. However, applying this scenario to high-frequency-peaked BL Lac objects (HBLs) with observed synchrotron peak frequencies of $\nu \sim 10^{17}$ Hz would require bulk Lorentz factors much higher than typical values of $\Gamma \sim 10$ inferred for blazars in general, unless the magnetic field in the CM frame is $\gg$ Gauss. If the jet composition is dominated by ions in both the spine and the sheath as assumed in this paper, one expects a small population of AGNs viewed under very small viewing angles with $\theta_{obs} \ll 1/\Gamma$, with very hard gamma-ray spectra (such as those detected by LAT onboard Fermi Observatory), while a larger population of off-axis AGNs with $\theta_{obs} > 1/\Gamma$ appear to have much softer spectra (cf. Fig.8). The recently emerging class of extreme BL Lac objects (e.g., Bonnoli et al. 2015) may possibly represent the small population of extremely narrowly beamed, very-hard-spectrum blazars expected in the ion-dominated shear flow scenario.

ACKNOWLEDGMENTS

This work was partially supported by NSF AST1313129 and Fermi Cycles 4 & 5 GI grants to Rice University, and NASA Fermi GI Grant no. NNX12AE31G to Ohio University. The work of M.B. is supported through the South African Research Chairs Initiative (SARChI) of the Department of Science and Technology and the National Research Foundation[1] of South Africa under SARChI Chair grant no. 64789. We thank Drs. Fabrizio Tavecchio and Gabriele Ghisellini for useful discussions. Simulations with the Zohar-II code were supported by the Lawrence Livermore National Laboratory.

Figure Captions

Fig.1 Setup of the (initially unmagnetized) shear flow PIC simulations of the e-ion plasma. This paper focuses on the longitudinal P-mode evolution in the x-y plane only, since the transverse T-mode saturates at very low level compared to the P-mode. In the present case the plasma consists of right-moving plasma in the central 50% of the y-grid, referred to as the "spine", sandwiched between left-moving plasmas at the top 25% and bottom 25% of the y-grid, referred to as the "sheath". The simulation box has periodic boundary conditions on all sides. Inset: Sketch illustrating d.c. magnetic field creation by the ECCI. <u>Throughout this paper and in all Figures, spatial scales are in units of the electron skin depth $c/\omega_e$.</u>

Fig.2 (a) Evolution of energy components of the 2D e-ion shear flow with $p_o=5$: EM field energy (A, black), electron energy (B, red), ion energy (C, blue). At late times the EM field energy saturates at ~12% of total energy; (b) Evolution of energy components of the 2D shear flow with $p_o=15$: EM field energy (A, black), electron energy (B, red), ion energy (C, blue). At late times the EM field energy also saturates at ~12% of total energy. However, the electron energy approaches the ion energy faster in the $p_o=15$ case than in the $p_o=5$ case.

Fig.3 Comparison of spatial profiles of (a) $\mathbf{B_z}$, (b) $\mathbf{E_y}$, (c) $\mathbf{E_x}$, (d) $\mathbf{j_x}$, (e) $\rho$ = net charge = $(n_i-n_e)$, between $p_o=5$ (left columns) and $p_o=15$ (right columns) cases at three different times: $t\omega_e$ = 1000 (top), 3000 (middle), 12000 (bottom). We note that the shear boundary layer thickness of the $p_o=5$ case is ~ half that of the $p_o=15$ case. Color scales show that the maximum $\mathbf{B_z}$, $\mathbf{E_y}$ fields of the $p_o=5$ case is much lower than those of the $p_o=15$ case.

Fig.4 Comparison of density profiles n vs. y (averaged over x) between the $p_o=5$ and $p_o=15$ cases at $t\omega_e = 8000$. Curve 1 (red) electrons, Curve 2 (blue) ions, Curve 3 (black) net charge ($=n_i-n_e$) These density profiles highlight the persistence of the ion vacuum gap, which is wider for the $p_o=15$ case than the $p_o=5$ case. Electrons, however, are not fully evacuated from the boundary layer, leading to the formation of the charge triple layer (double capacitor), which plays an important role in electron energization.

Fig.5 Comparison of electron distribution function $f_e(\gamma)$ (particle no. per unit $\gamma$) vs. $\gamma$ for (a) the $p_o=5$ case with that of (b) the $p_o=15$ case at $t\omega_e = 10000$; (c)(d) Same as (a)(b) for ion distribution function $f_i(\gamma)$ at $t\omega_e = 10000$. We see that the electron energy peaks at $\gamma_e \sim \gamma_i m_i/m_e$ in both cases.

Fig.6 Phase plot ($\mathbf{p_y}$ vs $\mathbf{p_{xLab}}$) of spine electrons for the $p_o=5$ case (a) compare to that of the $p_o=15$ case (b) at $t\omega_e=8000$, after Lorentz boosting $\mathbf{p_x}$ to the "laboratory frame" in which the sheath is initially at rest. By this time, some of the spine electrons have diffused into the sheath region and are decelerated, forming the low-energy bow-shaped population at left. The arrow-shaped high energy population corresponds to electrons remaining in the spine. Note that electrons are more concentrated at the highest energy for the $p_o=15$ case.

Fig.7 Phase plots (**y** vs. $\mathbf{p_{xLab}}$) of spine electrons for the $p_o=5$ case (a) compare to that of the $p_o=15$ case (b) at $t\omega_e=8000$, after Lorentz boosting $\mathbf{p_x}$ to the "laboratory frame" in which the sheath is initially at rest. By this time, some of the spine electrons have diffused into the sheath region and are decelerated, forming the low-energy population at left. The central high energy

population corresponds to electrons remaining in the spine. Note that electrons are more concentrated at the highest energy for the $p_o$=15 case.

Fig.8 Distribution of the tangent of the "beam angle" (=$|p_y/p_{xLab}|$) of spine electrons vs. Lorentz factor $\gamma_{Lab}$ in the "laboratory frame" at $t\omega_e$=8000 for (a) $p_o$=5 and (b) $p_o$=15. We see that in both cases all of the high-energy spine electrons (i.e. those that did not cross over to the sheath and get decelerated) have beam angles much smaller than 1/$\Gamma$ (red dashed lines). In both cases there exists an anti-correlation between beam angle and electron energy.

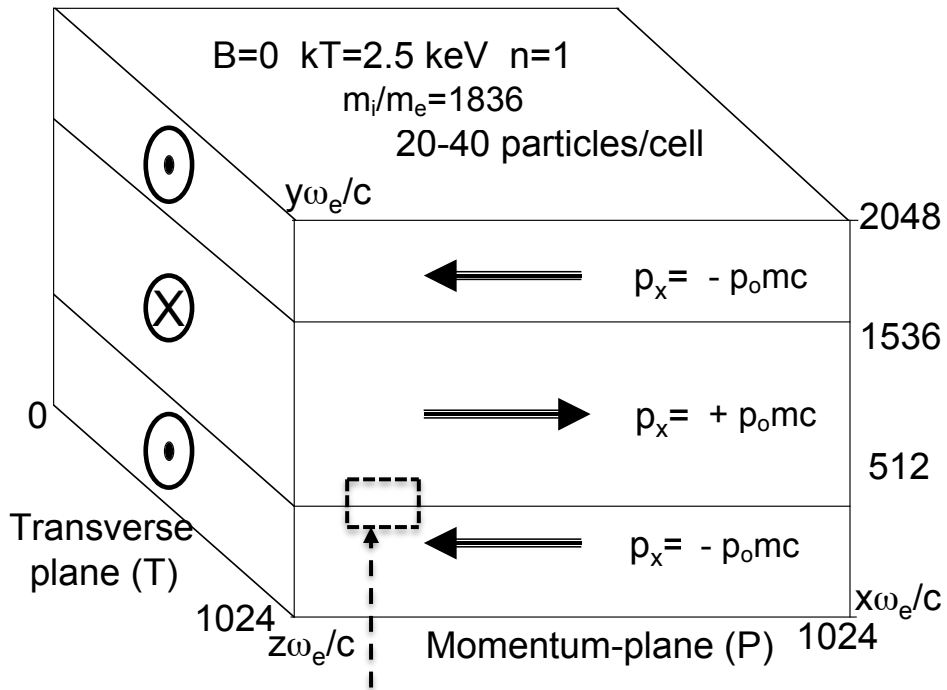

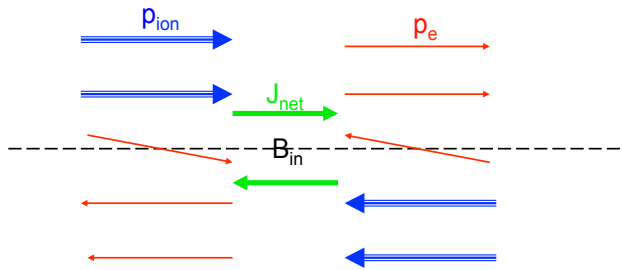

Fig.1

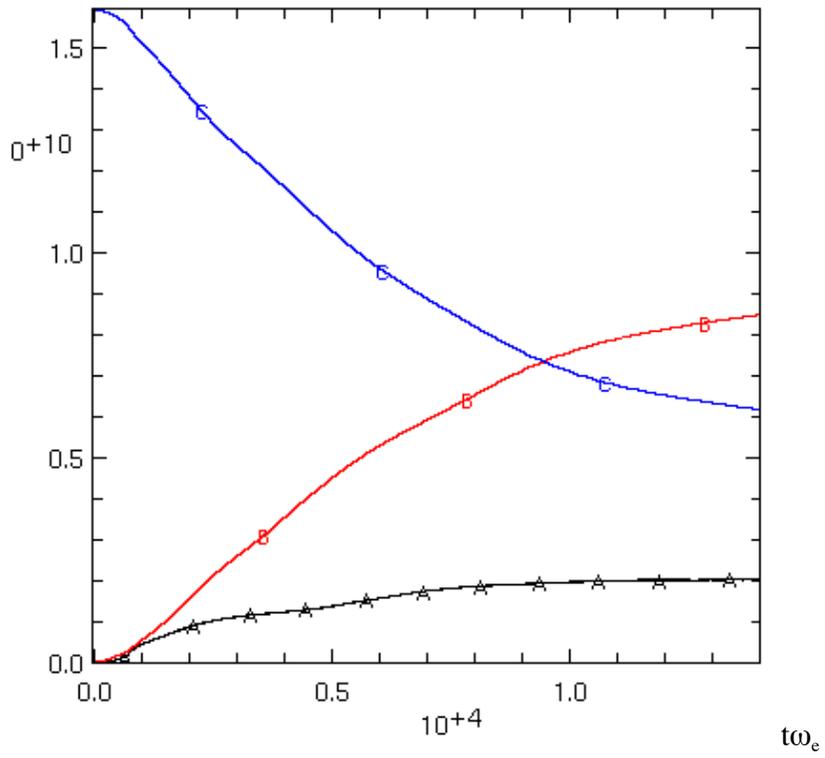

(a) $p_o=5$

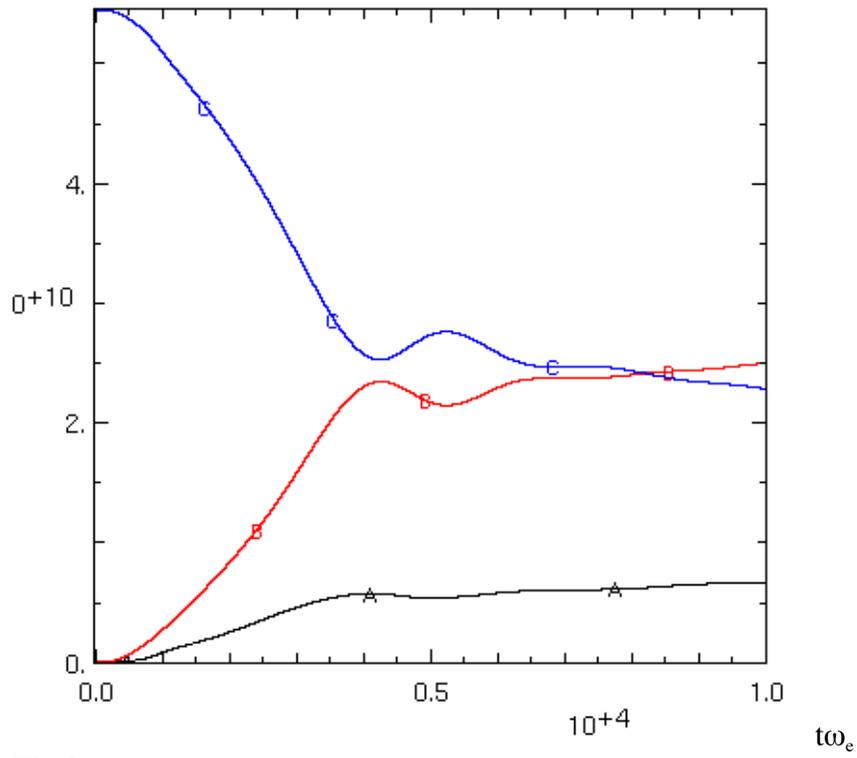

(b) $p_o=15$

Fig.2

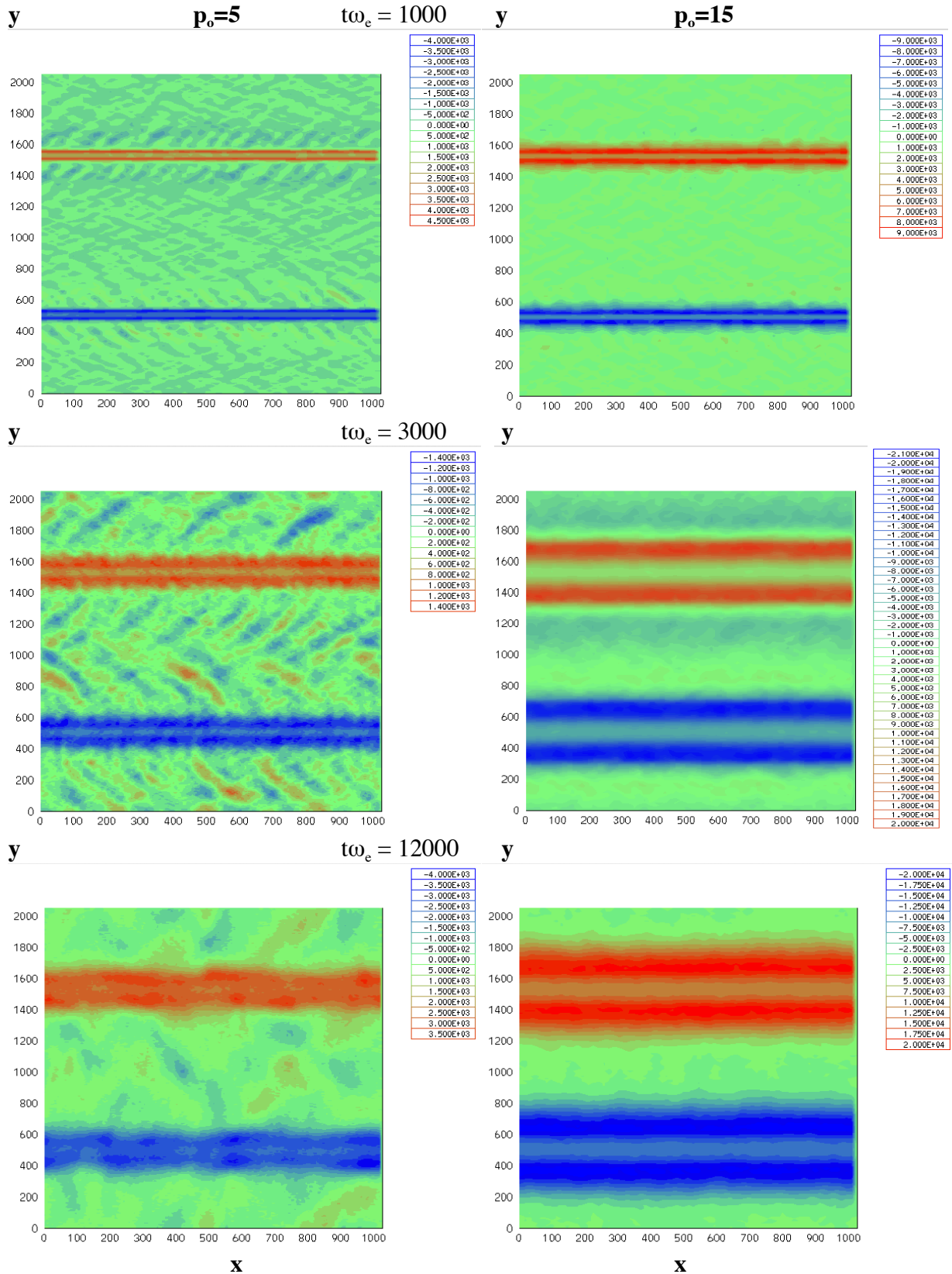

Fig.3(a) $B_z$

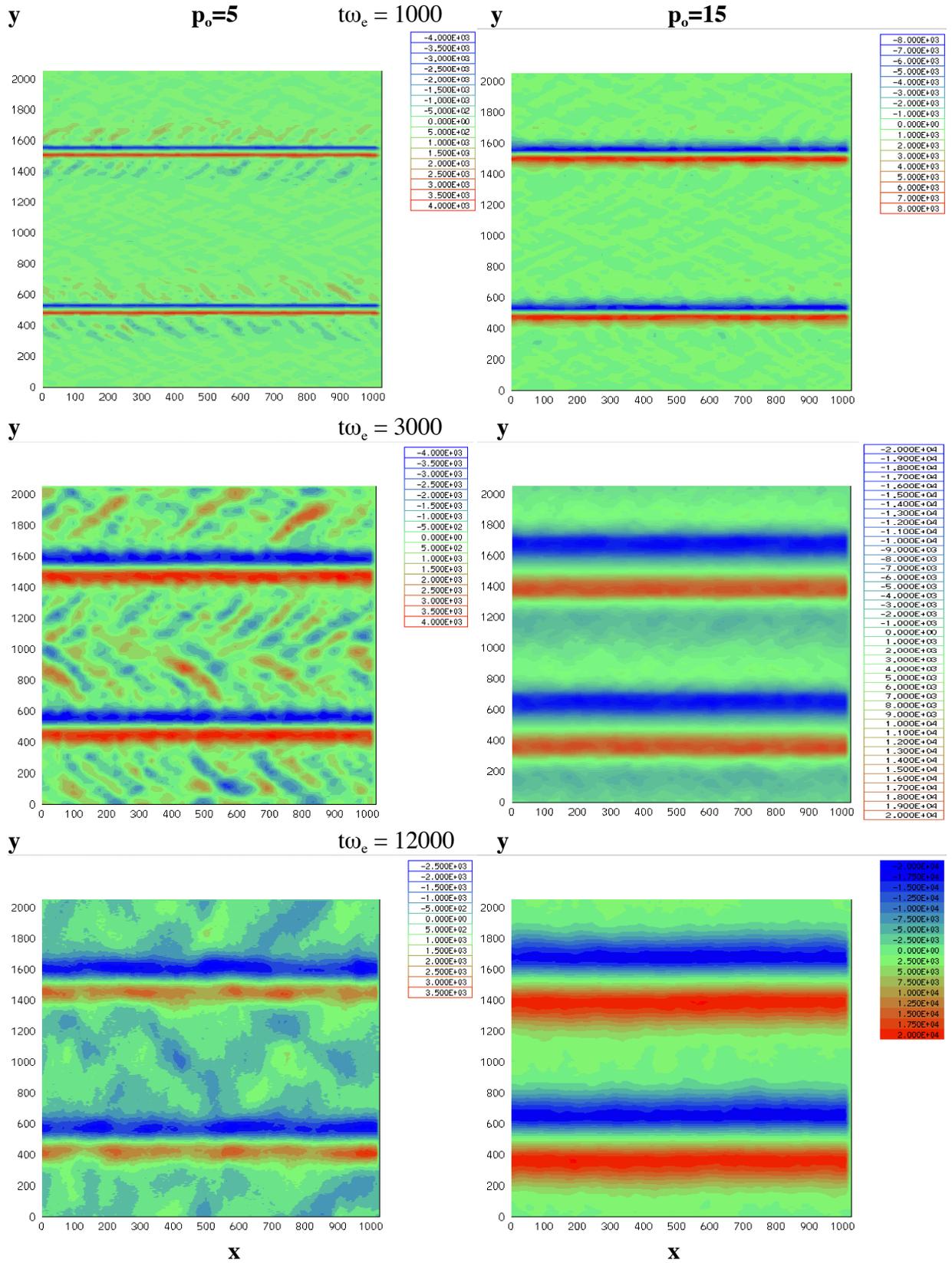

Fig.3(b) $E_y$

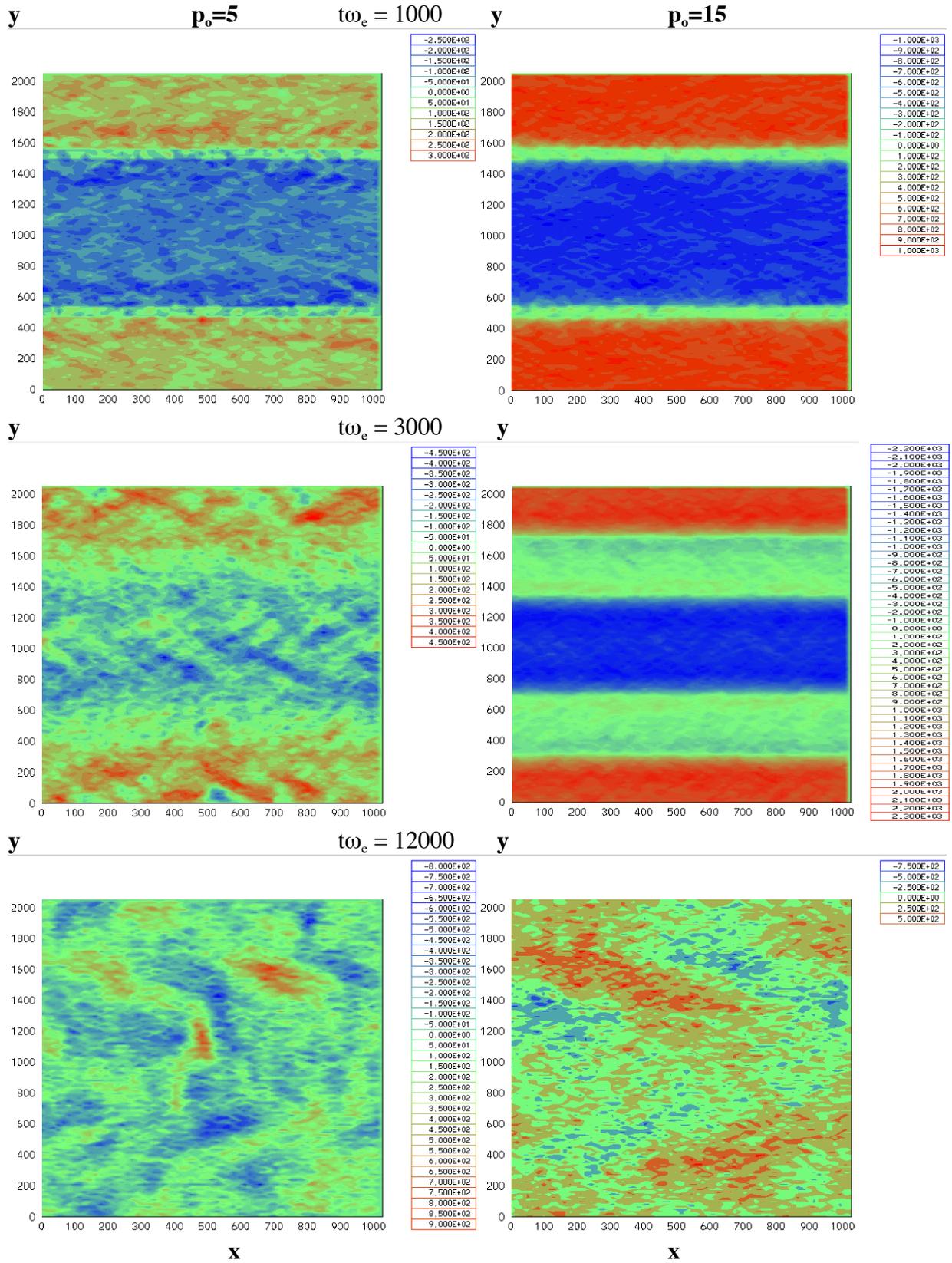

Fig.3(c) $E_x$

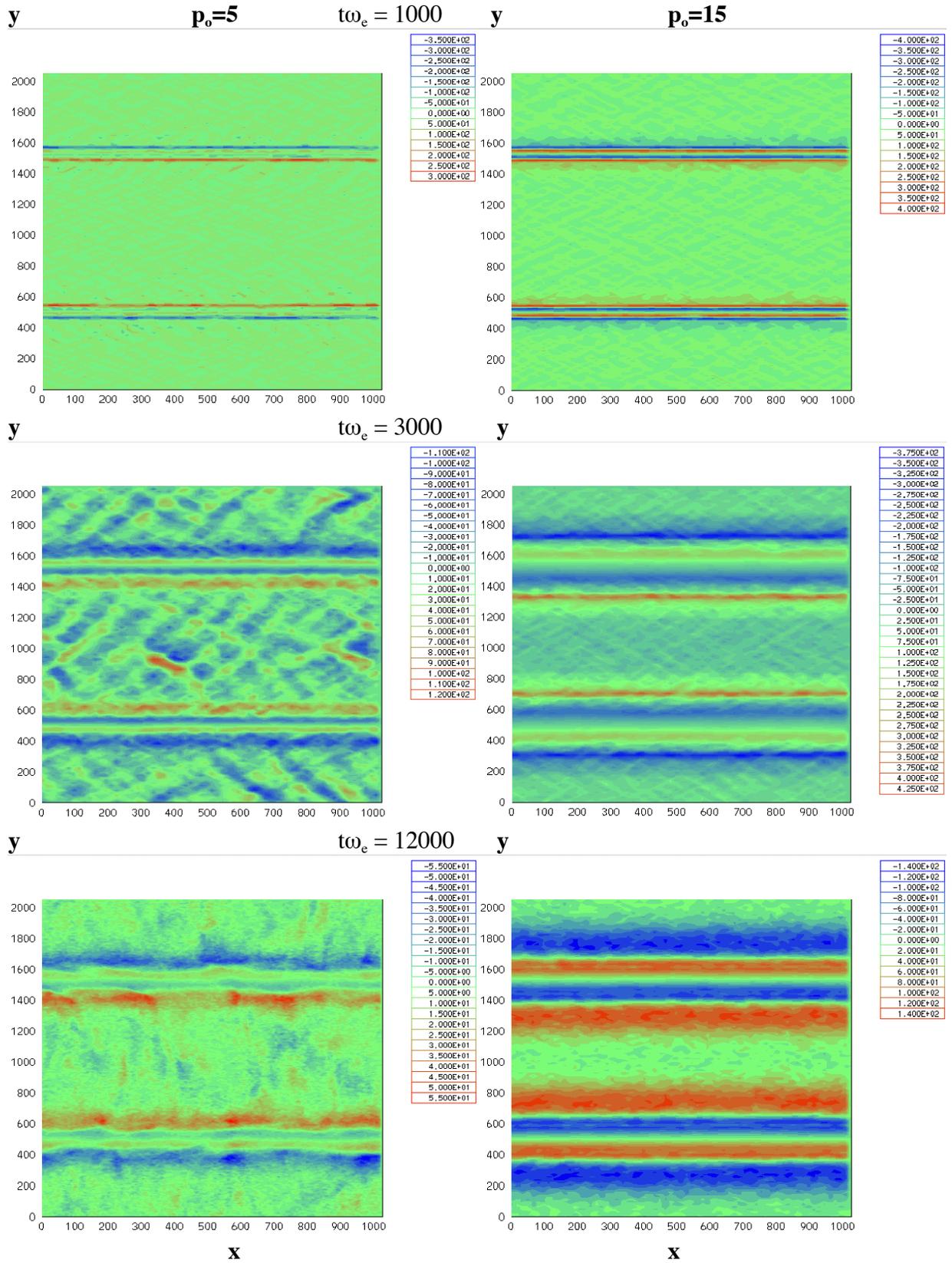

Fig.3(d) $j_x$

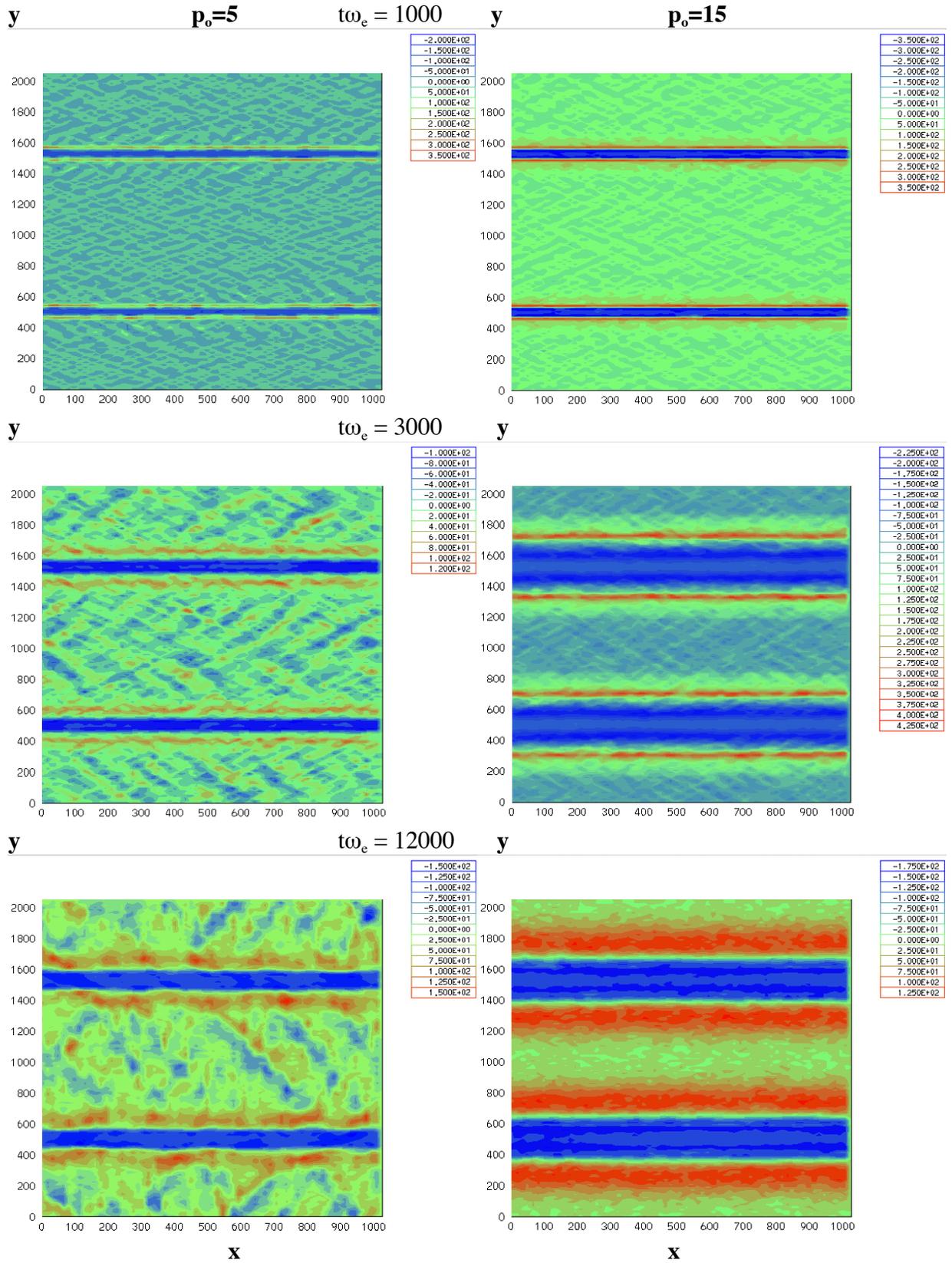

Fig.3(e) $\rho$

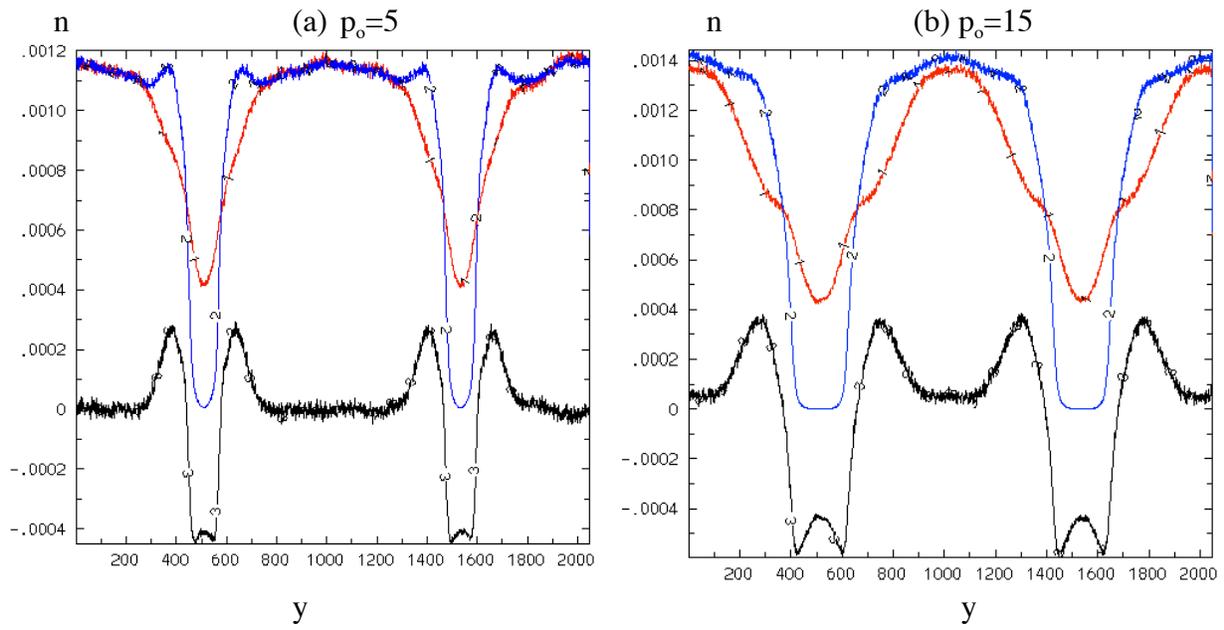

Fig.4

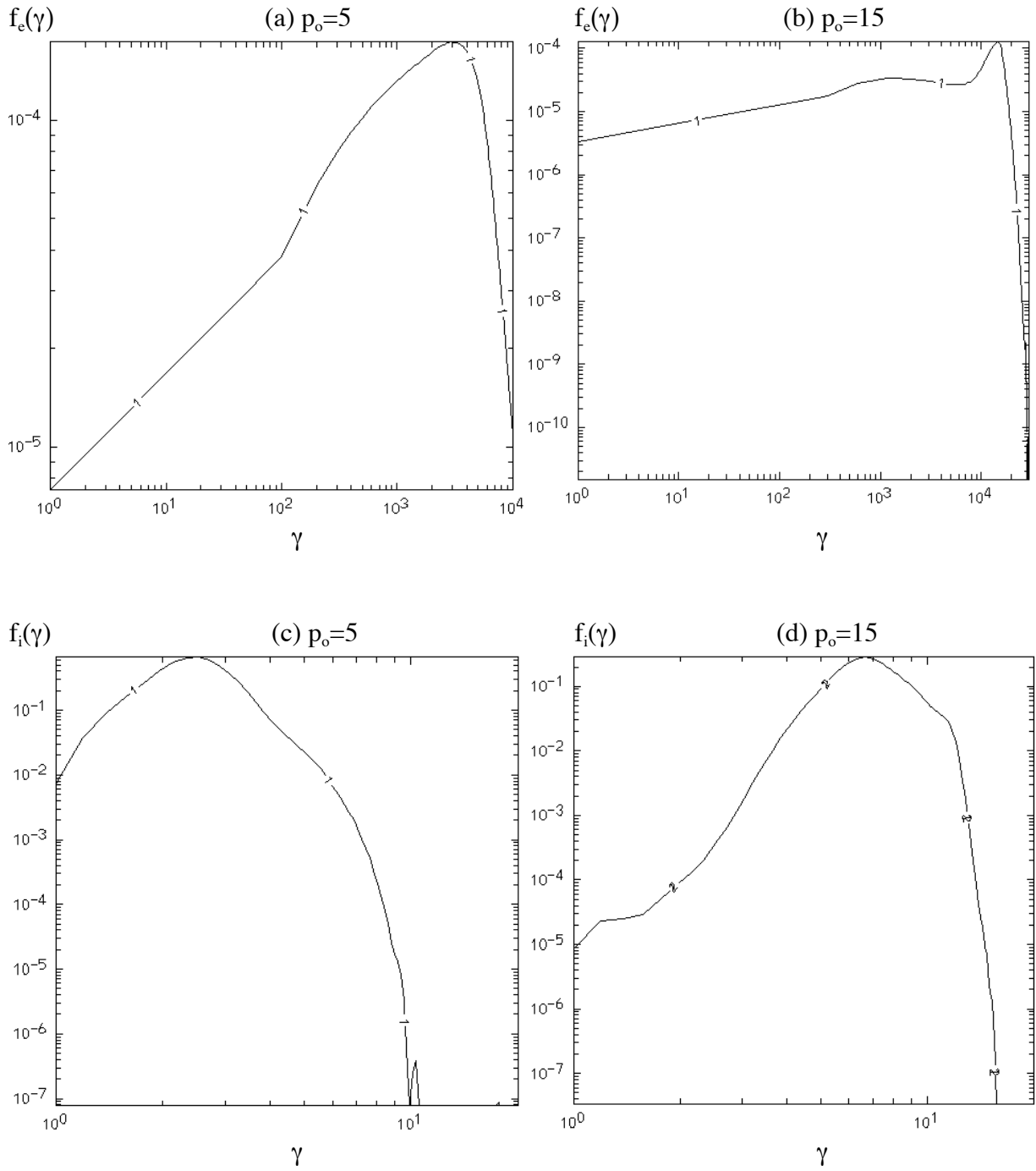

Fig.5

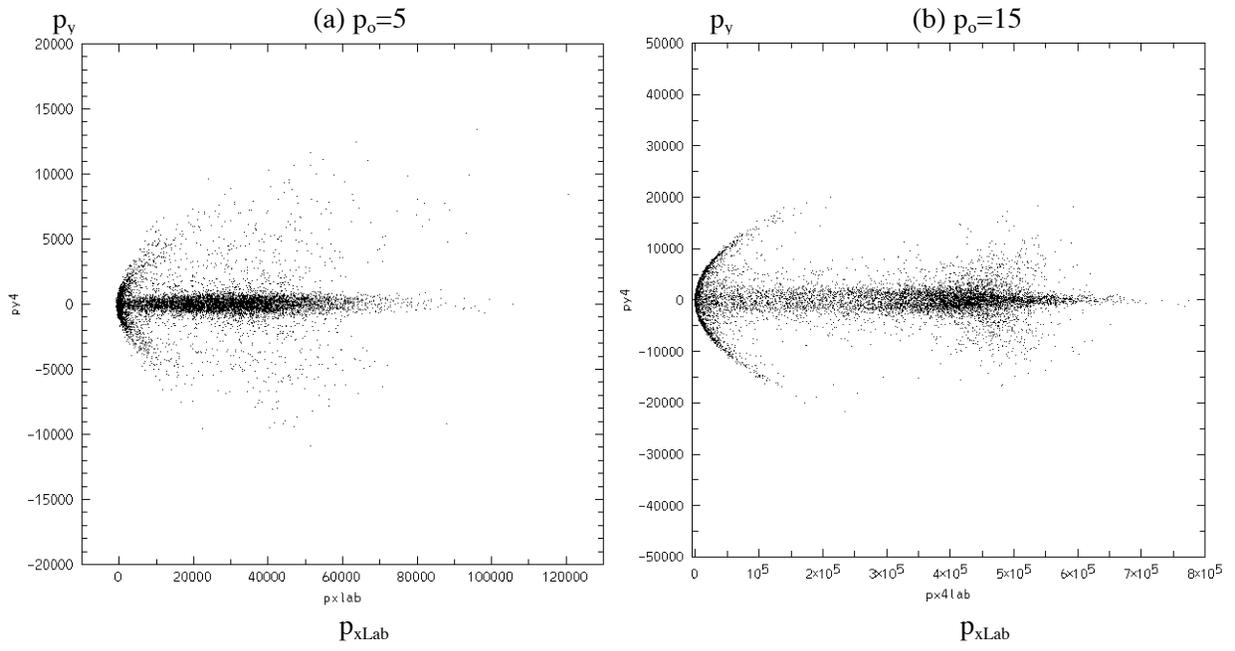

Fig.6

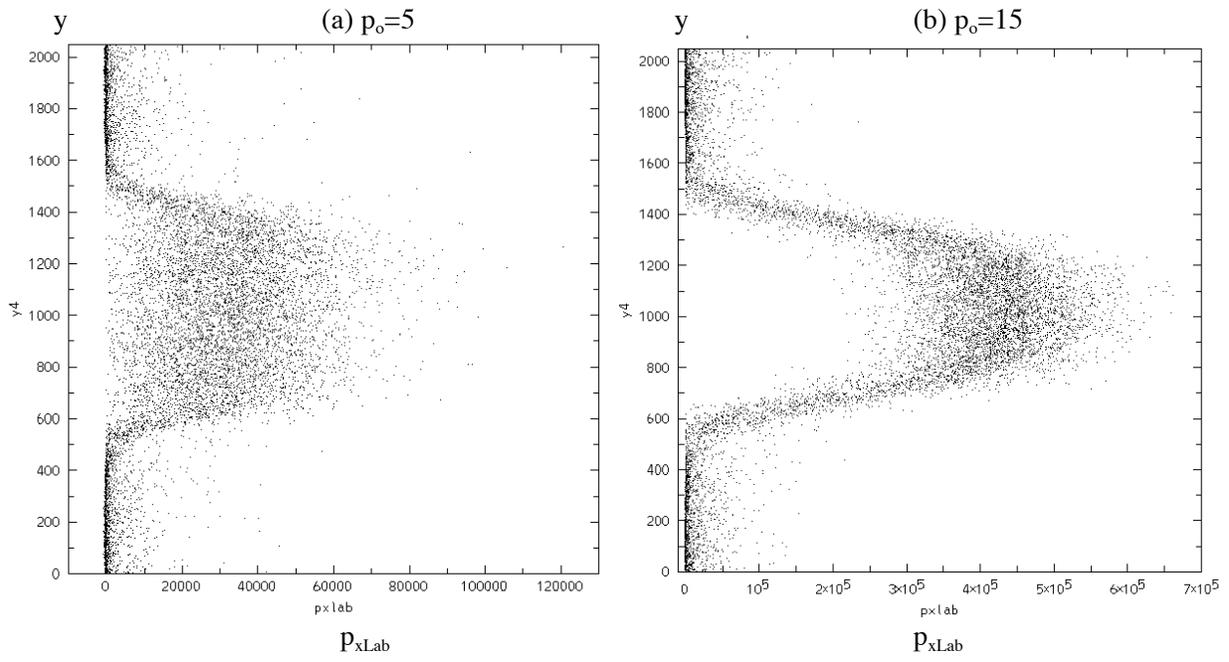

Fig.7

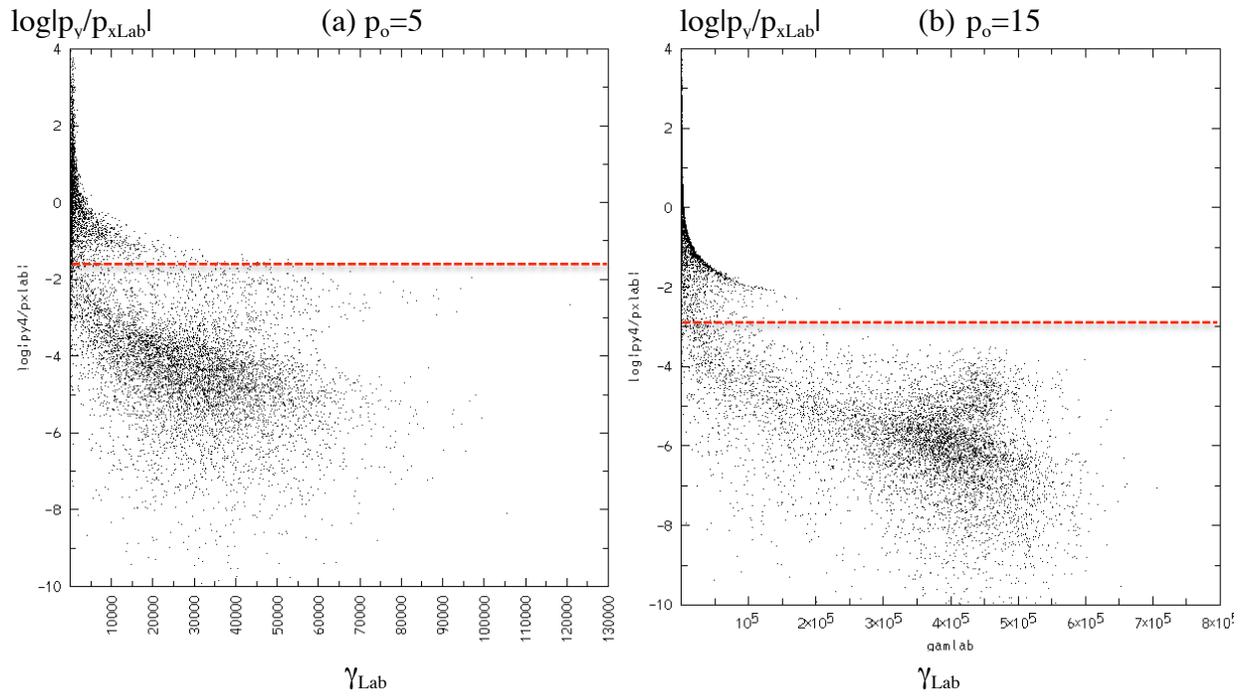

Fig.8